# Towards atomic-resolution quantum measurements with coherently-shaped free electrons

Ron Ruimy[†1], Alexey Gorlach[†1], Chen Mechel[1], Nicholas Rivera[2], and Ido Kaminer[1]

1. *Solid State Institute, Technion-Israel Institute of Technology, Haifa 32000, Israel*
2. *Department of Physics, Massachusetts Institute of Technology, Cambridge, MA, 02139, United States*
kaminer@technion.ac.il; † equal contributors

Free electrons provide a powerful tool to probe material properties at atomic-scale spatial resolution. Recent advances in ultrafast electron microscopy enable the manipulation of free electron wavefunctions using laser pulses. It would be of great importance if one could combine the spatial resolution of electron probes with the ability of laser pulses to probe coherent phenomena in quantum systems. To this end, we propose a novel technique that leverages free electrons that are coherently-shaped by laser pulses to measure quantum coherence in materials. Developing a quantum theory of electron-qubit interactions in materials, we show how the energy spectrum of laser-shaped electrons enables measuring the qubit Block-sphere state and decoherence time ($T_2$). Finally, we present how such shaped electrons can detect and quantify superradiance from multiple qubits. Our scheme could be implemented in an ultrafast transmission electron microscope (UTEM), opening the way towards the full characterization of the state of quantum systems at atomic-scale resolution.

**I. Introduction**

Electron microscopy and spectroscopy are powerful methods to extract information about quantum emitters such as atoms, molecules, and solids [1]. State-of-the-art techniques include cathodoluminescence (CL) [1-3] and electron energy-loss spectroscopy (EELS) [1, 4, 5], which measure excitation energies, bandgaps, and the local density of photonic states (LDOS) [6] with high spatial resolution. However, there is a lot of information that cannot normally be extracted, such as the quantum state of material excitations, decoherence times ($T_2$), and more generally, any information associated with off-diagonal components of the density matrix of quantum systems. Looking beyond electron microscopy, it is of fundamental interest to find what quantum information is exchanged in the interaction of free electrons and a general quantum emitter. Henceforth, for brevity, we refer to the quantum systems as "qubits", since we focus on a single material excitation that corresponds to a fixed electron transition.

For qubits in the optical range, their quantum state and decoherence times can be analyzed using laser-based coherent control [7] and pump-probe laser experiments [8- 13]. Yet, probing individual qubits in optical experiments are limited in spatial resolution by the optical wavelength: measuring an individual emitter, rather than an ensemble, requires a dilute sample. Dilute ensembles cause the acquisition of sufficient signal to become a substantial challenge. Often, having high densities of qubits is intrinsic to their fabrication, and important for applications [14-16], such as semiconductor quantum dot (SCQD) devices for quantum science and technology [17-19] or applications involving qubit-qubit interactions for quantum gates and quantum information processing [20-22]. Therefore, being able to measure decoherence rates of *dense* emitter ensembles is highly-desired; for example, to differentiate between $T_2$ and $T_2^*$ [23] and also to distinguish between coherent and incoherent broadening in emitters [24]. Such a distinction is also of technological relevance, e.g., in high color-contrast displays [25]. In light of these challenges, there are great prospects for using free electrons as highly localized probes that could quantify quantum features of individual qubits – including decoherence rates and the full quantum state.

Here we propose a scheme to access the coherent quantum information of qubits using coherently-shaped free electrons. We show how controlling the incident shaped electron enables measuring the qubit state, and extracting the longitudinal ($T_1$) and transverse ($T_2$) relaxation rates of the qubit. Such measurements can be done using a pump-probe scheme in which the pump is a laser pulse and the probe is the electron. We develop a theory to describe the interaction of such a shaped electron with a two-level system characterized by a transition dipole moment, relevant to many systems (e.g., defect centers, direct band-gap transitions) and emblematic of many other types of excitations in materials. The results are especially strong in cases where the dipole is large, as with excitons in 2D semiconductor heterostructures [26], perovskites [27-29], and Rydberg state atoms [30, 31].

Our work is motivated by recent advances in ultrafast electron spectroscopy and microscopy [1], especially photon-induced nearfield electron microscopy (PINEM) [32-35] that showed how free electrons can be coherently-modulated using femtosecond-pulsed lasers.

Our work suggests taking advantage of such shaped electrons and of the fundamental electron–qubit interaction to extract the qubit state. We quantify how this concept can be implemented in electron microscopes using EELS and CL and how it can enhance their signal. These enhancements match part of the semiclassical predictions in [36], based on classical modulation of free electrons. We show how the quantum description goes beyond this theory and provides capabilities that cannot be modeled semiclassically.

Apart from measuring $T_2$, our scheme can also measure $T_1$, which provides information about the LDOS. Importantly, $T_1$ can also be extracted at high spatial resolution using recent advances in time-resolved CL [1, 2]. In comparison, our scheme could potentially provide a better temporal resolution – the PINEM femtosecond timescales. Such short timescales enable to probe phenomena such as a superradiance [37, 38] in regions where the density of qubits is high enough to satisfy the superradiance condition. Our shaped electron schemes could lead to a full toolbox of high-resolution capabilities for reading/writing arbitrary states of qubits in materials.

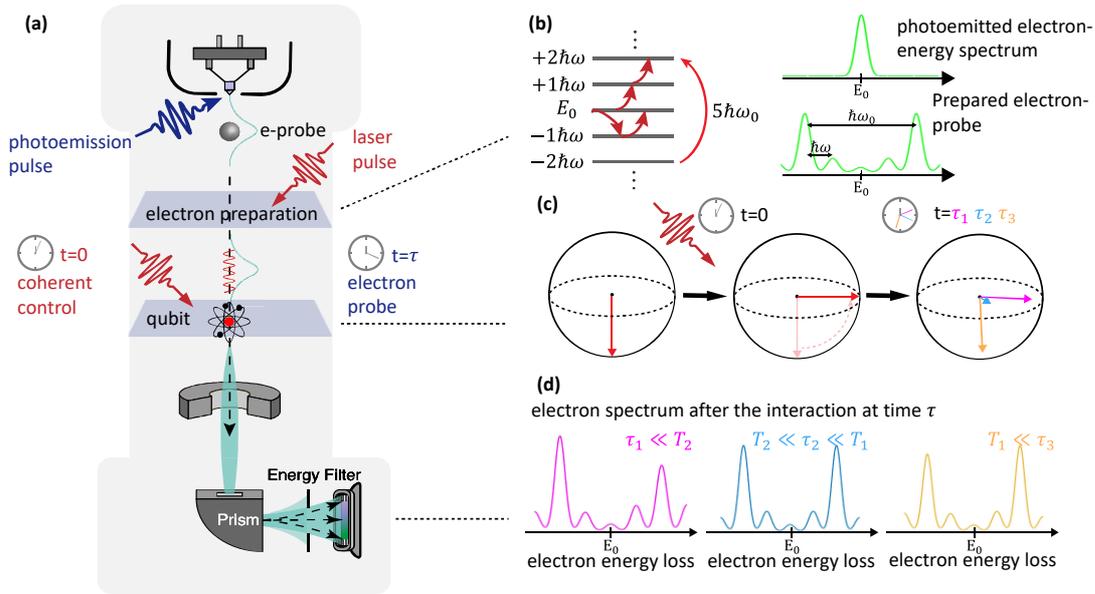

**Fig.1 Coherently-shaped free electrons as high-resolution probes of coherence in quantum systems: the measurement scheme.** (**a**) An ultrafast transmission electron microscope (UTEM) can create and then (**b**) coherently-shape an electron probe via the PINEM interaction. Following this interaction, the shaped electron is a superposition of energies shifted by integer multiples of a driving laser frequency $\omega$. We choose $\omega$ such that two energy peaks will be separated by the qubit energy gap $\hbar\omega_0$ (in this case $g^{\text{PINEM}} = 1.5$). (**c**) Another laser excites the qubit to the state $\frac{1}{\sqrt{2}}|g\rangle + \frac{i}{\sqrt{2}}|e\rangle$; after time delay $\tau$, the shaped electron interacts with it. (**d**) The interaction is measured using electron energy loss spectroscopy (EELS), measured as a function of delay $\tau$. This scheme enables us to measure different properties of the qubit. For example, at small $\tau$ we get a strong asymmetry in the electron spectrum, which disappear for $\tau$ larger than $T_2$. For $\tau$ larger than $T_1$, the qubit goes through relaxation before it interacts with the electron, thus the EELS only changes by a small increase to the loss peaks.

## II. Quantum interaction of qubits and free electrons

We can model the interaction of a free electron and a qubit by a Hamiltonian of the form:

$$H = -i\hbar v \partial_z + \frac{\hbar \omega_0}{2}\sigma_z + \frac{e}{4\pi\varepsilon_0} \cdot \frac{(\boldsymbol{d}_\perp \cdot \boldsymbol{r}_\perp + d_\| z)\sigma_+ + (\boldsymbol{d}_\perp^* \cdot \boldsymbol{r}_\perp + d_\|^* z)\sigma_-}{(r_\perp^2 + z^2)^{3/2}}, \qquad (1)$$

where the first term describes the Hamiltonian of the free electron and the second term describes the Hamiltonian of the qubit. The average speed of the electron wavepacket is $v$. The qubit is characterized by energy separation $\hbar\omega_0$ between the ground $|g\rangle$ and excited $|e\rangle$ states; $\sigma_\pm, \sigma_z$ are the standard Pauli matrices. The third term in Eq. (1) describes the interaction with the transition dipole moment $\boldsymbol{d} = \langle g|e\boldsymbol{r}|e\rangle$, with components $d_\| = d_z$ and $\boldsymbol{d}_\perp = \hat{\boldsymbol{x}} d_x + \hat{\boldsymbol{y}} d_y$. $r_\perp$ is the minimum distance between the center of the electron wavepacket and the center of the qubit (the impact parameter), and $\varepsilon_0$ is the vacuum permittivity. The full derivation of Eq. (1) is in Supplementary Material (SM) Section II.

The two main approximations behind Eq. (1) are: (I) The paraxial approximation for the electron since the electron energy is much larger than that of the excitation. (II) External decoherence channels by other material excitations [39] such as Bremsstrahlung radiation [40] and characteristic X-Ray [41] occur at probability much smaller than unity; thus, the reduced density matrix of the electron and the qubit (for the excitation of interest) will be the same as if the external channels were not considered at all. See SM Section I for further discussion.

The scattering matrix can be found from the Magnus expansion [42] as:

$$S = e^{-i(g\sigma^+ b + g^* b^+ \sigma^- + \kappa \sigma_z)}. \qquad (2)$$

Here, the operators $b$ and $b^+$ are momentum translation operators for the electron, $b = e^{i\omega_0 z/v}$ [43] (in the limit of electron energy much larger than $\hbar\omega_0$, they equivalently describe energy translation); $g$ and $\kappa$ are the interaction parameters.

The interaction strength $g$ is typically small ($|g| \ll 1$) and then, can be approximated as:

$$g = \frac{ed_x\omega_0 K_1\left(\frac{\omega_0 r_\perp}{v}\right)}{2\pi\varepsilon_0 \hbar v^2} + \frac{ed_z\omega_0 K_0\left(\frac{\omega_0 r_\perp}{v}\right)}{2\pi\varepsilon_0 \hbar v^2}, \tag{3}$$

where $K_0(x)$ and $K_1(x)$ are modified Bessel functions of the second kind; a more general derivation for relativistic electron–qubit interactions are described in SM Section III, showing corrections to $g$ (that do not change the concepts shown below). The interaction constant $\kappa$ in Eq. (2) under the approximation of $|g| \ll 1$ can be neglected:

$$S = e^{-i(g\sigma^+ b + g^* b^+ \sigma^-)} = \cos|g| - i \cdot \sin|g|\left(e^{i\phi_g}\sigma_+ b + e^{-i\phi_g}\sigma_- b^+\right), \tag{4}$$

where $\phi_g$ is the phase of the interaction constant $g$.

Consider a qubit prepared (e.g., by an optical pulse) in a coherent superposition of its excited $|e\rangle$ and ground $|g\rangle$ states $|\psi\rangle = a_1|g\rangle + a_2|e\rangle$. The qubit undergoes decoherence and relaxation (Fig. 1c), leading to a density matrix at time $\tau$ (SM Section IV):

$$\rho_a(\tau) = \begin{pmatrix} 1 - |a_2|^2 e^{-\frac{\tau}{T_1}} & e^{-\frac{\tau}{T_2}} a_1 a_2^* e^{i\omega_0 \tau} \\ e^{-\frac{\tau}{T_2}} a_1^* a_2 e^{-i\omega_0 \tau} & |a_2|^2 e^{-\frac{\tau}{T_1}} \end{pmatrix}, \tag{5}$$

A shaped electron interacts with this qubit. The density matrix after the electron–qubit interaction is $\rho_f = S^+ \rho_i S$, where $\rho_i = \rho_a \otimes \rho_e$ is the initial density matrix of the joint system and $\rho_e$ is the initial density matrix of the shaped electron. The electron density matrix after the interaction $\mathrm{tr}_{\mathrm{qubit}}\, \rho_f$ contains information about the qubit state. We measure the EELS of the post-interaction electron, which provides the diagonal of the electron density matrix (Fig. 1d). This measurement contains information about the off-diagonal terms of Eq. (5) if the initial electron is a superposition of two (or more) energies that differ by the qubit energy gap $\hbar\omega_0$. The generation of such a shaped electron is possible via the PINEM interaction (Fig. 1a and Fig. 1b).

**III. Measurement of relaxation and decoherence times**

We first consider the interaction of a conventional electron that was not pre-shaped by the laser – known in electron microscopy as the zero-loss peak and called below "unshaped electron". The EELS features of an unshaped electron interacting with a qubit are proportional to $|g|^2$. For example, a pump-probe scheme with unshaped electrons can measure $T_1$ (SM, Section V), as we show in Fig. 2a. Since in realistic scenarios the coupling constant $|g| \ll 1$ (a typical value is $|g| \sim 10^{-3}$), the qubit features could be very hard to see.

This problem can be resolved by shaped electron, for which the interaction creates EELS features that are proportional to $|g|$ instead of $|g|^2$. Moreover, there are qubit properties that remain inaccessible for an unshaped electron. One such property is $T_2$. To probe $T_2$, we take a superposition electron state of two (or more) different energies (Fig. 1b). Such shaped electrons lead to interference in the electron–qubit interaction, depending on the phases: $\phi_g$ of the coupling constant, $\phi_e$ between the two electron energies (determined by the modulating laser), and $\phi_a$ between the excited and ground qubit states. We denote the phase difference by $\Phi = \phi_a - \phi_e - \phi_g$. All phases are considered in the rotating frame of reference.

The EELS probabilities of electron energy gain $P_+$ and energy loss $P_-$ depend on the phase difference $\Phi$. Their difference $\Delta P$, to first order in the coupling strength $|g|$, is:

$$\Delta P = P_+ - P_- = (P_+ + P_-)|g|e^{-\frac{\tau}{T_2}}\sin(\Phi). \qquad (6)$$

The maximal EELS signal is given by choosing $\Phi = \pi/2 + \pi\mathbb{Z}$. This can be done without explicit knowledge of $\phi_{a,e}$ simply by scanning different electron phases until maximal contrast is reached. Repeating this experiment for different $\tau$ allows measuring the decoherence time $T_2$ (Fig. 2b).

A remarkable quality of Eq. (6) is that interference enables to get a net energy gain or loss proportional to $|g|$ rather than $|g|^2$, which increases the sensitivity of EELS. This quality can be exploited in multiple ways. For example, we can propose a sensitive $T_1$ measurement scheme by applying a $\pi/2$ pulse right before the interaction with the shaped electron, and then a similar scheme to the one measuring $T_2$ can extract $T_1$.

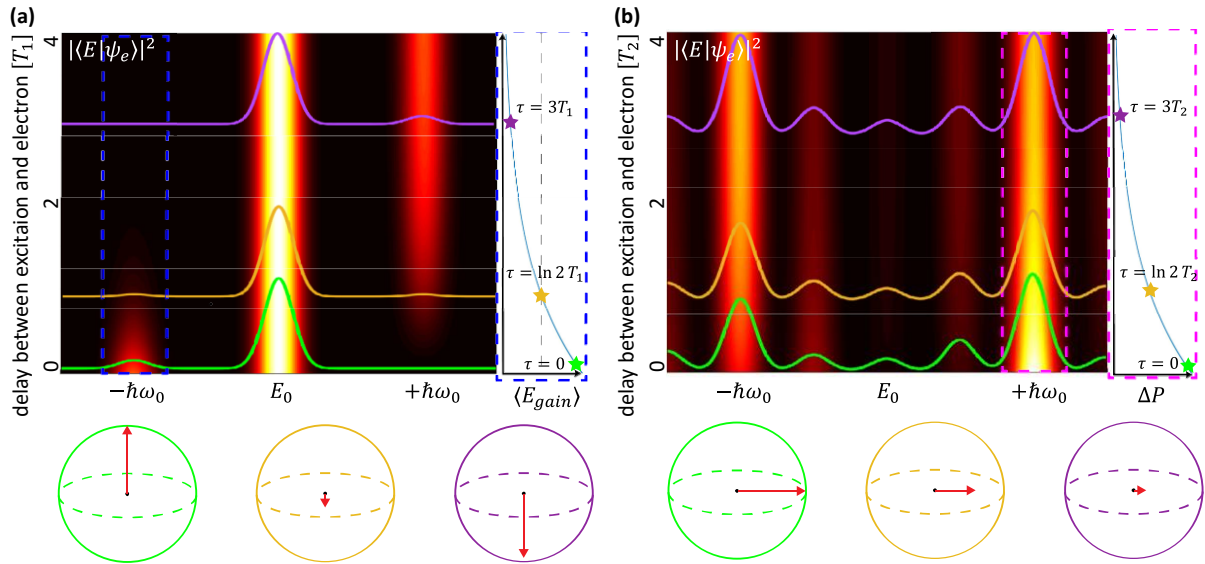

**Fig.2 Extracting the relaxation ($T_1$) and decoherence ($T_2$) times from a sequence of delayed electron energy spectra.** (a) A laser pulse excites the qubit to $|e\rangle$, which then relaxes back with a characteristic time $T_1$. Unshaped electron probes interact with the qubit with delays $\tau$, and are measured using electron energy loss spectroscopy (EELS). $T_1$ is extracted from the sequence of EELS spectra. $\langle E_{\text{gain}}\rangle$ is the average energy gain for an unshaped electron after the interaction. (b) A laser pulse excites the qubit to $\frac{1}{\sqrt{2}}(|g\rangle + i|e\rangle)$, which then processes while decohering with the characteristic time $T_2$. Shaped electron probes interact with the qubit with delays $\tau$, and are measured using EELS. $T_2$ is extracted from the sequence of EELS spectra. To emphasize the concept, we used $g = 0.3$ for the plots. $\Delta P$ is the difference between the EELS probabilities of energy gain of $\hbar\omega_0$ and energy loss of $\hbar\omega_0$.

**IV. Finding the exact qubit state on the Bloch-Sphere**

Finding the qubit state from its interaction with a shaped electron could be of particular interest for setting the initial condition or measuring the final state in a quantum simulator. So far, the use of optical tools for controlling individual qubits motivated setting larger-than single wavelength distances between qubits, which can only be done in certain experimental platforms, and limits qubit-qubit interactions. Here, the prospects of shaped electron probes are in enabling new experimental platforms as candidate quantum simulators, facilitated by free electron's high spatial resolution.

For these reasons, we present a scheme that utilizes shaped electrons (Fig. 3a) for measuring the qubit state on the Bloch-sphere. Consider a general qubit state $|\psi_a\rangle = \cos\left(\frac{\theta_a}{2}\right)|e\rangle + e^{i\phi_a}\sin\left(\frac{\theta_a}{2}\right)|g\rangle$, with $\theta_a$ and $\phi_a$ representing the location on the Bloch-sphere in spherical coordinates. The qubit interacts with the shaped electron $|\psi_e\rangle = |E_+\rangle + e^{i\phi_e}|E_-\rangle$ (Fig. 3b). The resulting EELS probability difference between the electron energy gain $P_+$ and energy loss $P_-$, to first order in $|g|$, is given by (illustrated in Fig. 3c)

$$\Delta P = (P_+ + P_-)|g|\sin(\theta_a)\sin(\Phi), \qquad (7)$$

which is elaborated in SM Section V and Section VIII. The quantity of $\sin(\theta_a)$ can be found by repeating the measurement multiple time, scanning over a range of relative phases $\phi_e$ until the probability difference is maximized, the probability difference in this point will be $g \cdot \sin(\theta_a)$.

The same measurement scheme can be used to extract additional information of interest, such as the size of the transition dipole moment (by extracting $g$ from Eq. (3)) and measuring the phase of the electron $\phi_e$ (by changing the qubit phase $\phi_a$ using laser excitations with different phases).

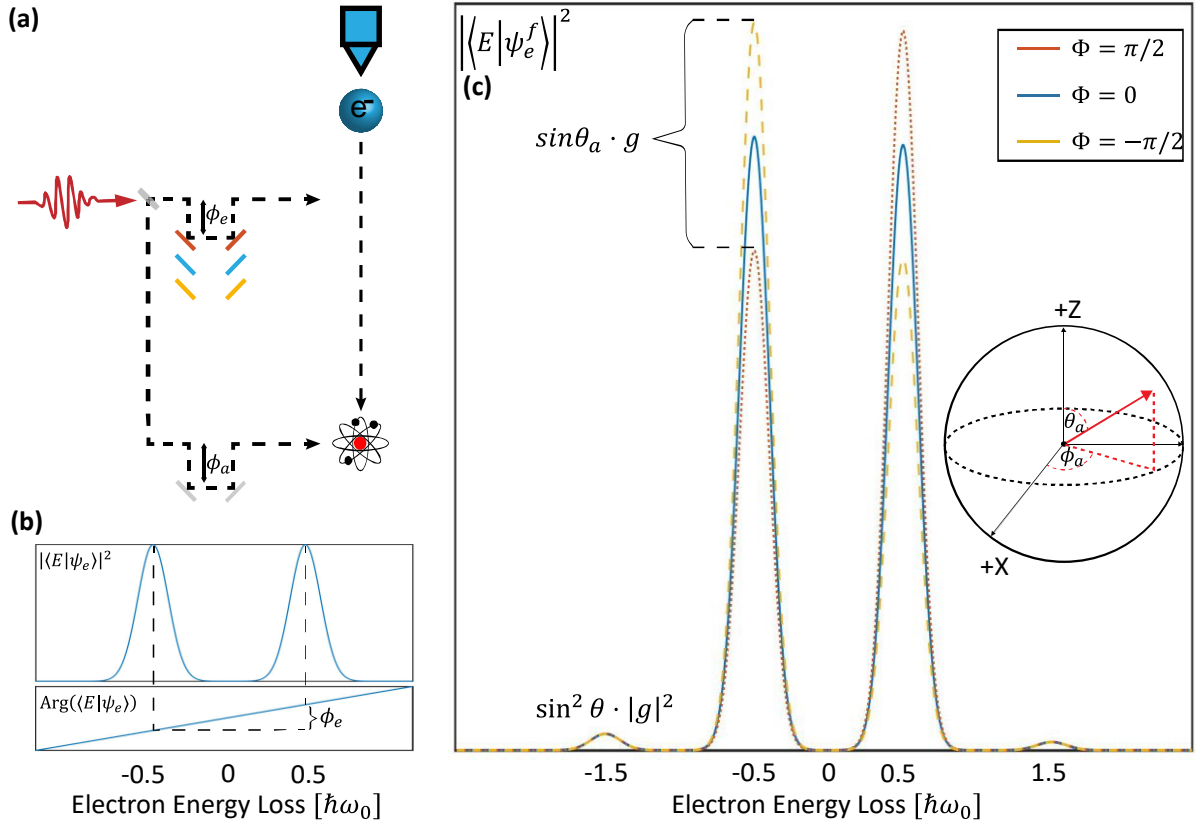

**Fig.3 A scheme for measuring the qubit state on the Bloch sphere:** **(a)** The electron is coherently-shaped by a laser pulse with a tunable optical path length that changes the relative phase $\phi_e$ between the electron's energy states. The same laser (or another that is phase-locked) excites the qubit to a state defined by angles $\theta_a$ and $\phi_a$ on the Bloch sphere. **(b)** The electron energy superposition state with relative phase $\phi_e$. **(c)** The resulting EELS spectra for different $\phi_e$. The difference between the maximal and minimal EELS peaks extracts the qubit state on the Bloch sphere. $\phi_g$ is the phase of the interaction constant $g$ and $\Phi = \phi_a - \phi_e - \phi_g$.

## V. Probing superradiance from multiple emitters

In cases of qubits with weak non-radiative relaxation, $T_1$ is related to the qubit's spontaneous emission rate $\gamma = 1/T_1$, which is influenced by the optical environment through the local density of photonic states (LDOS). Thus, the shaped-electron–qubit interaction provides a new way to measure the LDOS through a pump-probe scheme – as elaborated in SM Section VII. Such a measurement scheme is shown in Figs. 4a-b.

This measurement scheme has temporal resolution on femtosecond timescales [32-35, 44, 45], gradually reaching attosecond timescales [35, 46- 48]. Consequently, the scheme can be utilized to explore novel emission phenomena on short timescales, such as superradiance (Fig.

4c) that occurs when a few qubits are bunched together. In superradiance, several excited qubits have a faster joint emission rate $\gamma$ than the $\gamma$ of each individual qubit. Our scheme can observe the rapid decay via changes in the EELS peaks. In this way, we can quantify the superradiant decay of even just a small number of emitters (SM, Section VI). The superradiance implies an enhancement in CL experiments (SM Section VI). A different CL enhancement can be created even for a single qubit by using shaped free electrons, as discussed further in SM Section V. The latter corresponds to part of the predictions of [36], based on a semiclassical theory.

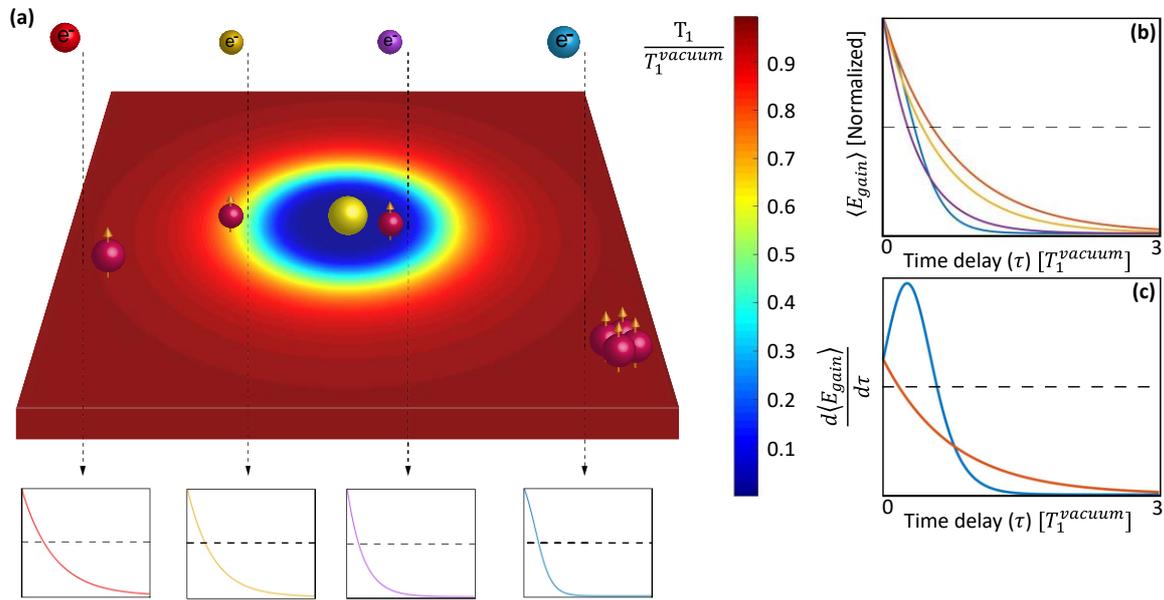

**Fig.4 Free-electron–qubit interactions for studying superradiance and mapping the local density of photonic states (LDOS). (a)** We exemplify the concept by placing qubits on a membrane surrounding a 10nm Au nanoparticle, which varies the LDOS and the resulting $T_1$ as a function of distance. Bottom schematic plots: the decay of the average electron energy gain $\langle E_{\text{gain}} \rangle$ as a function of delay time $\tau$, for different qubit positions. **(b)** Comparison between normalized $\langle E_{\text{gain}} \rangle$ plots for different qubits. **(c)** Extracting features of superradiance from the derivative of normalized $\langle E_{\text{gain}} \rangle$, which also represents the emitted power. Orange: a single qubit. Blue: multiple qubits; showing the typical superradiance emission peak with an increased rate of emission at a short (nonzero) time.

## VI. Discussion and outlook

The main challenge limiting implementations of the above concept is the typically small $|g|$. Consequently, the most attractive quantum systems for a proof-of-concept are ones with a large transition dipole moment such as perovskite nanocrystals [27-29]. Besides the requirement for a large dipole moment, the interaction strength also depends on the electron velocity $v$ and the impact parameter $r_\perp$. Eq. (3) shows that $g$ increases for small $r_\perp$ and small $v$ (so long as the electron is not so slow that the paraxial approximation breaks). The optimal electron velocity for a dipole that is transverse to the electron velocity is $v \approx \frac{r_\perp \omega_0}{1.33}$. Substituting typical parameters for perovskite nanocrystals with transition dipole moment $\approx$ 288 D (6 nm) and energy gap of $\sim$ 3eV [27-29], and setting $r_\perp$ to be equal to 6nm, we find an optimal electron velocity $v/c \approx$ 7%, corresponding to 1.25 keV acceleration voltage (SM Section V).

The resulting coupling parameter is $|g| \approx 0.1$, which can be readily detected in typical EELS detectors. Such interaction conditions enable a temporal resolution of about a hundred femtoseconds and a spatial resolution that can reach 1.4 nm at these electron energies [49]. The energy resolution is limited by the bandwidth of the excitation laser (~10 meV and below [45, 50]). With such a large $|g|$, finding the state of the qubit (or any other property) with a relative error of 1% requires a few million electrons (with current laser repetition rates, this implies integration time of ~1 s [45]). Note that standard EELS systems and current PINEM experiments are performed at higher velocities (e.g., 80 keV electrons, show an optimal coupling parameter $|g|$ = 0.025, well within current detection capabilities). Nevertheless, highly sensitive EELS also exist at low acceleration voltages [51].

We note that for $r_\perp$ smaller than the size of the dipole, there could be beyond-dipole corrections to the electron–qubit interaction. Then, the qubit wavefunction profile can no longer be ignored. When the size of the quantum dot can be larger than the size of the electron probe, as with SCQDs, we may be able to probe the qubit state and even decoherence time as a function of position inside a quantum dot. Such experiments can give insight into the effects of many-body physics in quantum emitters.

Having electron spot sizes of few nanometers open intriguing possibilities for testing dense ensembles of qubits in systems such as SCQDs [17-19], and measuring the decoherence rates and other quantum properties of individual quantum dots in the ensemble. The combined temporal and spatial resolutions can help research currently unobserved phenomena such as fluctuations of $T_2$ in the ensemble, as a function of the distance between qubits, and even as a function of time. Generally, the electron–qubit interaction can probe the mechanisms differentiating between $T_2$ and $T_2^*$ [23].

High-density SCQDs are of particular interest in areas of quantum technologies, being able to demonstrate fundamental coherent quantum effects such as Rabi oscillations [18, 52], Ramsey fringes, photon echoes, and even quantum coherent revival [53-55]. Currently, these phenomena are probed optically, which limits the ability to study the properties of single qubits out of a dense ensemble and understand how they vary within the ensemble.

We should also discuss the validity of modeling real materials with a two-level system (qubit) for the interaction with free electrons. Unlike a photon, which typically only creates a single excitation, an electron will typically set off many excitations in a given material – these additional excitations act as decoherence channels, which become entangled to the electron. When the additional excitations are separated in energy from the qubit energy, then one can post-select electrons (using an energy filter) that loses or gains the qubit energy, ensuring that the measured electrons interact with the qubit. For thin samples where the electron only weakly interacts with the decoherence channels, the qubit-electron density matrix is unchanged by these channels. Despite this, future work should further consider whether these decoherence channels alter the information available in the electron and our ability to infer the qubit state. Looking down the line, finding what material systems realize the idealized electron–qubit interaction will be a key question in this field.

To conclude, we predict that coherently-shaped free electrons can determine the qubit state, its decoherence time, and other quantum characteristics. We envision the combination of this idea with proposals of coherent control using shaped electrons [36, 56]: Together, they constitute the building blocks for using shaped electrons to read and write the quantum state of atoms, molecules, and quantum dots. Such capabilities, especially if achieved at deep subwavelength and potentially even atomic resolutions, are attractive for creating new kinds of quantum simulators. We envision quantum simulators in which shaped electrons enable depicting the initial state of each element of a quantum system, and enable reading the final (or intermediate) states, using femtosecond (and eventually attosecond [35, 46-48, 57]) time resolution.

---

We note that a few days before submission, we have become aware of another work done in parallel to our work that partially overlaps with it [58]. There, the interaction of the shaped electron with qubits was also treated fully-quantum mechanically, and it was also proposed that such interactions can enhance the electron energy gain/loss spectrum and probe the qubit coherence.

# Towards atomic-resolution quantum measurements with coherently-shaped free electrons

## Supplementary Material

Ron Ruimy[†1], Alexey Gorlach[†1], Chen Mechel[1], Nicholas Rivera[2], and Ido Kaminer[1]

[1] *Solid State Institute, Technion-Israel Institute of Technology, Haifa 32000, Israel*
[2] *Department of Physics, Massachusetts Institute of Technology, Cambridge, MA, 02139, United States*

kaminer@technion.ac.il; † equal contributors

**Section I – Validity of the two-level system (TLS) model**

In our work, we are interested in the possibility of using free electrons as a probe for quantum coherence in qubit systems. The measurements are done through electron energy loss spectroscopy (EELS) and we model the interaction as an interaction between the evanescent electric field of the free electron and the dipole moment of the TLS. This model predicts that the electron can gain/lose a single quantum of energy while de-exciting/exciting the quantum TLS. In reality, modeling an emitter as a TLS is often not a good description, as a free electron beam scattering from atomic electrons can lose energy through a wide variety of phenomena. While an electron can excite a particular transition of interest through the mechanism outlined above, an electron can excite many additional transitions (such as core-shell transitions). Beyond the atom undergoing many types of transitions, the electron can also lose energy by many different channels besides collision with the emitter of interest. The electron can lose energy via Bremsstrahlung processes, as well as via collective excitations (photons, phonons, and plasmons).

All these inelastic processes occur at probability much smaller than unity for thin samples. The majority of the probability remains in electrons that do not undergo inelastic processes, also called the zero-loss peak (ZLP) in EELS. We are interested in EELS peaks that are related to a particular transition in the system (the "qubit" or "two-level-system" transition). We consider energy loss associated with other peaks in the EELS spectrum as decoherence channels. Thus, in practice, whenever we measure an electron (after passing through the sample) with energy different than $E_0$ or $E_0 \pm \hbar\omega_0$, we ignore it. We only consider materials for which the qubit transition is separated in energy than the other inelastic processes.

One concern is that different EELS peaks are not necessarily well-separated, especially if some are broad (such as the peaks associated with plasmons). Those broad peaks may overlap the peak fitting the TLS transition. This problem is partially solved using electrons in a superposition of two (or more) energy states, all spaced by $\hbar\omega_0$. In this case, what we measure is not the height of the peaks related to the TLS transition but the height difference between the two ZLPs (corresponding to the two initial energies of the electron) as they interfere with each other while interacting with the TLS. This interference between two initial energies rely on the coherence of the qubit and can significantly enhance the contrast in the measurements, as analyzed in the paper. Competing excitations such as plasmon and phonon polaritons have very short coherence lifetime and so will not lead to the resulting interference between the two electron energies, i.e., will result in a comparably small change to the EELS spectra. Thus, we can neglect the decoherence channels, if the energy difference between the two ZLPs matches the qubit energy, and assuming that the probabilities of the decoherence channels is much smaller than unity (not necessarily much smaller than the original spontaneous qubit transition). This idea is illustrated in Fig. 1.

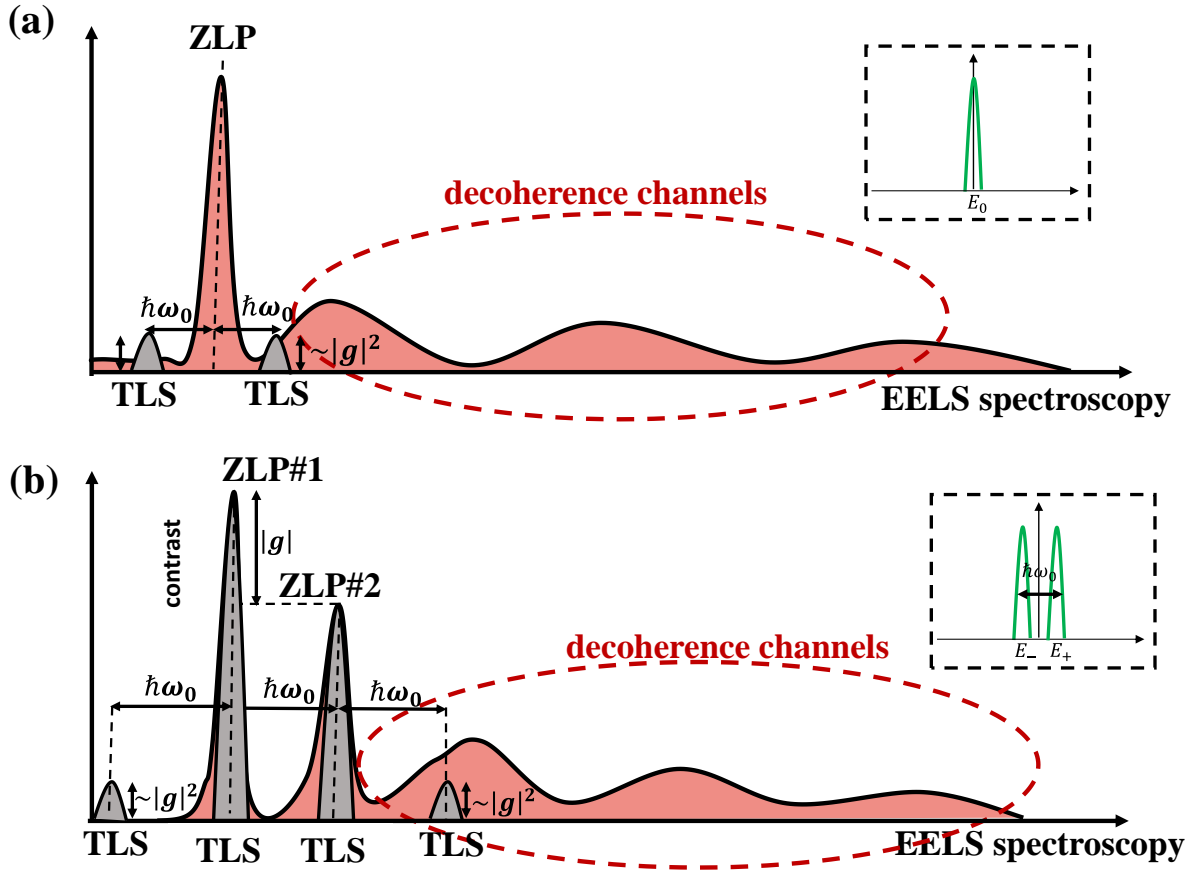

**Fig 1. Schematic electron energy-loss spectrum (EELS) that highlights the electron transitions that can be controlled by coherently-shaped free electrons**. (**a**) The spectrum of a unshaped electron (electron wavefunction with Gaussian distribution of energy): the peak of the two-level system (TLS) excitation scales like $|g|^2$ and can often be much smaller than other processes that we call "decoherence channels". (**b**) The electron is shaped to have two different energies that can be thought of as two different zero-loss peaks (ZLPs). Their relative phase before the interaction creates a contrast in their height after the interaction that contains information about the state of the qubit. The contrast scales like $|g|$ if the qubit is in a coherent superposition state. The key to this enhancement is the interference of the shaped electron. Other contributions to the spectrum (here noted as decoherence channels), such as surface plasmon losses or peaks due to core-loss transitions [1], can be neglected when their probability is small enough so that they do not significantly alter the ZLP(s).

**Section II – The quantum theory for the interaction of a free electron and a TLS**

In this section, we discuss the interaction of a free electron with a quantum TLS through the dipole interaction in a fully quantum mechanical manner. We develop the 1D Hamiltonian that is used in order to get the results in the main text. This 1D Hamiltonian is based on the paraxial approximation for the electron, which is justified in the limit where the electron initial energy is much higher than the energy of the TLS.

The Hamiltonian consists of the free electron Hamiltonian ($H_e$), the TLS Hamiltonian ($H_{TLS}$), and the interaction Hamiltonian ($V$)

$$H = H_0 + V = H_e + H_{\text{TLS}} + V. \tag{1}$$

For a TLS with an energy gap of $\hbar\omega_0$ its Hamiltonian is represented by the Pauli matrix $\sigma_z$ so that $H_{TLS} = \frac{\hbar\omega_0}{2}\sigma_z$ with $\sigma_z = |e\rangle\langle e| - |g\rangle\langle g|$, with $e$ the excited state and $g$ the ground state. The interaction part of the Hamiltonian couples the TLS dipole moment to the electric field associated with a moving point charge (electron) $V = -\boldsymbol{d} \cdot \boldsymbol{E}$. This interaction term is justified under the dipole approximation in which we assume that the bound electron wave function is tightly confined to the atomic nuclei. The transition dipole operator is defined as follows:

$$\boldsymbol{d} = \langle e|e\boldsymbol{r}|g\rangle \cdot |e\rangle\langle g| + \langle g|e\boldsymbol{r}|e\rangle \cdot |g\rangle\langle e| = (\boldsymbol{d}_{eg}\sigma_+ + \boldsymbol{d}_{eg}^*\sigma_-), \tag{2}$$

where $\boldsymbol{d}_{eg} = d_x\hat{\boldsymbol{x}} + d_y\hat{\boldsymbol{y}} + d_z\hat{\boldsymbol{z}}$. In this notation, we assume that the transition dipole operator is purely off-diagonal which typically occurs due to the inversion symmetry of atoms. For a relativistic (spin-less) electron, the free electron Hamiltonian is given by Klein-Gordon [2],

$$H_e = c\sqrt{m^2c^2 + \boldsymbol{p}^2}. \tag{3}$$

We now simplify the Klein-Gordon Hamiltonian under the case in which we treat the electron under the paraxial approximation, which results from linearizing the dispersion relation of the electron around its central momentum. Put more rigorously, we will restrict the electron Hamiltonian to the space of functions of the form $\psi = e^{i k_0 \cdot r} f$, where f is slowly varying ($|\nabla f| \ll |kf|$). In that case, the action of the electron Hamiltonian on the state is:

$$H_e \psi = c\sqrt{m^2 c^2 - \hbar^2 \nabla^2} e^{i k_0 \cdot r} f \approx e^{i k_0 \cdot r} c \sqrt{m^2 c^2 + \hbar^2 k_0^2 - 2 i \hbar^2 \boldsymbol{k_0} \cdot \nabla f}. \tag{4}$$

Taylor expanding and noting that in relativity $\frac{p}{E} = \frac{v}{c^2}$, we have:

$$e^{i \vec{k_0} \cdot \vec{r}} \left( c\sqrt{m^2 c^2 + \hbar^2 \boldsymbol{k_0}^2} + \frac{\hbar c \boldsymbol{k_0} \cdot (-i \hbar \nabla)}{\sqrt{m^2 c^2 + \hbar^2 \boldsymbol{k_0}^2}} \right) f = e^{i k_0 \cdot r} (E_0 - i \hbar \boldsymbol{v} \cdot \nabla) f \tag{5}$$

We can replace the Hamiltonian with $H_e = -i\hbar \boldsymbol{v} \cdot \nabla$ as they are the same in this space of states up to the identity. Physically, this approximation is saying that regardless of the electron's momentum, its velocity is v, which is a type of "no-recoil approximation". We will assume that the electron is moving on the z-axis and that the TLS is located at (0,0,0). We can then write the Hamiltonian as:

$$H = -i\hbar v \partial_z + \frac{\hbar \omega_0}{2} \sigma_z + \left(\boldsymbol{d}_{eg} \sigma_+ + \boldsymbol{d}_{eg}^* \sigma_-\right) \cdot \boldsymbol{E}(r_\perp, 0, z), \tag{6}$$

where $\boldsymbol{E}(r_\perp, 0, z)$ is the electric field at the location of the TLS (0,0,0) as a function of the electron's position $(r_\perp, 0, z)$ as given by the Maxwell equations. The motion of the electron and the position of the TLS are shown in Fig. 2

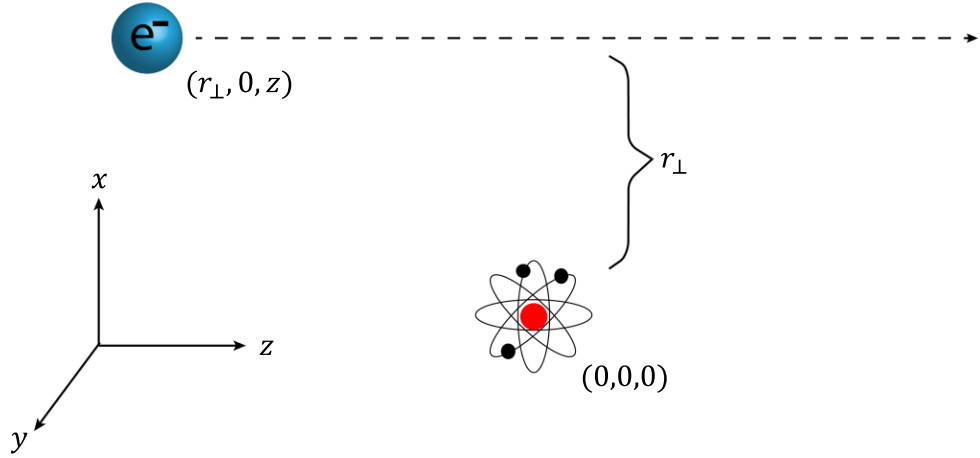

**Fig. 2. Schematic of the interaction between a free electron and an atom.** The electron has coordinates $x = r_\perp$, $y = 0$ and moves along the z-axis with speed $v$ so that $z = vt$. The atom is taken as stationary and situated at the origin $(0,0,0)$.

**Section III – Derivation of the scattering matrix and analysis of the interaction strength**

In this section, we find an approximate solution to the scattering matrix resulting from the Hamiltonian in Eq. (6) through the Magnus expansion. We discuss the structure of the complete scattering matrix and show that an approximate form of it is highly accurate under the most realistic conditions in electron-TLS interactions. To solve the problem, it will be convenient to move into the interaction picture in which:

$$V_I(t) = e^{i\frac{H_0}{\hbar}t} V e^{-i\frac{H_0}{\hbar}t}, \qquad (7)$$

where $H_0 = H_e + H_{TLS}$. Using the identity, $e^{vt\partial_z} f(z) e^{-vt\partial_z} = f(z + vt)$, we find:

$$V_I(t) = \left(\boldsymbol{d}\sigma_+ e^{i\omega_0 t} + \boldsymbol{d}^*\sigma_- e^{-i\omega_0 t}\right) \cdot \boldsymbol{E}(r_\perp, 0, z + vt). \qquad (8)$$

The S-matrix, $S = Te^{-\frac{i}{\hbar}\int_{-\infty}^{\infty} V_I(t)dt}$ (with T the usual time-ordering operator) can now be evaluated according to the Magnus expansion [3]:

$$S = Te^{-\frac{i}{\hbar}\int_{-\infty}^{\infty} V_I(t)\, dt} = e^{\sum_{k=1}^{\infty}\Omega_k(\infty)}, \tag{9}$$

where the Magnus expansion operators $\Omega_k$ are given by nested commutators of the $V_I$ at different times. For example, the first two terms of the expansion are given as:

$$\begin{cases} \Omega_1(\infty) = -\frac{i}{\hbar}\int_{-\infty}^{\infty} dt\, V_I(t), \\ \Omega_2(\infty) = \frac{1}{2}\left(-\frac{i}{\hbar}\right)^2 \int_{-\infty}^{\infty}\int_{-\infty}^{t_1} dt_1 dt_2\, [V_I(t_1), V_I(t_2)], \\ \cdots \end{cases} \tag{10}$$

where higher orders $\Omega_n$ can be calculated in the same way. The first order is completely analytical, but the general structure can be analyzed by understanding the form of the Magnus operators. For the first order, we will write the integral and perform the exchange of variables $x = z + vt$.

$$\Omega_1(\infty) = -\frac{i}{\hbar v}\int_{-\infty}^{\infty} dx\, \left(\boldsymbol{d}_{eg}\sigma_+ e^{\frac{i\omega_0}{v}(x-z)} + \boldsymbol{d}_{eg}^*\sigma_- e^{-\frac{i\omega_0}{v}(x-z)}\right) \cdot \boldsymbol{E}(r_\perp, 0, x). \tag{11}$$

To simplify this further, we now define the electron momentum ladder operators:

$$b = e^{-\frac{i\omega_0}{v}z},\quad b^+ = e^{\frac{i\omega_0}{v}z}, \tag{12}$$

which are operators that lower (or raise) the electron momentum by $\frac{\hbar\omega_0}{v}$. In the regime where the kinetic energy of the electron is significantly larger than the energy gap of the TLS, this translates to an energy lowering (raising) operator for the electron wave function that changes the electron's energy by that of the TLS ($\hbar\omega_0$). With this definition in place, $\Omega_1$ is given as:

$$\Omega_1(\infty) = -\frac{i}{\hbar v}\left(b\sigma_+ \boldsymbol{d}_{eg} \cdot F\{\boldsymbol{E}(r_\perp, 0, x)\}\left(\frac{\omega_0}{v}\right) + b^+\sigma_- \boldsymbol{d}_{eg}^* \cdot F^*\{\boldsymbol{E}(r_\perp, 0, x)\}\left(\frac{\omega_0}{v}\right)\right), \tag{13}$$

where $F$ denoting the Fourier transform. The Fourier components of the electric field of a relativistic point charge have a well-known expression and so we can find $\Omega_1$ exactly as [4]:

$$\Omega_1(\infty) = -\frac{ie\omega_0}{2\pi\varepsilon_0\gamma\hbar v^2}\left(d_x K_1\left(\frac{\omega_0 r_\perp}{v\gamma}\right) + id_z K_0\left(\frac{\omega_0 r_\perp}{v\gamma}\right)\frac{1}{\gamma}\right)(b\sigma_+) -$$

$$-\frac{ie\omega_0}{2\pi\varepsilon_0\gamma\hbar v^2}\left(d_x K_1\left(\frac{\omega_0 r_\perp}{v\gamma}\right) - id_z K_0\left(\frac{\omega_0 r_\perp}{v\gamma}\right)\frac{1}{\gamma}\right)(b^+\sigma_-), \tag{14}$$

where $K_m$ are modified Bessel functions of the second kind and $\gamma = \frac{1}{\sqrt{1-\beta^2}}$, with $\beta = \frac{v}{c}$. To the first order in the Magnus expansion, we can write the scattering matrix as:

$$S = e^{-i(gb\sigma_+ + g^*b\sigma_-)}, \tag{15}$$

where the interaction strength parameter $g$ given by:

$$g = \frac{e\omega_0}{2\pi\varepsilon\gamma\hbar v^2}\left(d_x K_1\left(\frac{\omega_0 r_\perp}{v\gamma}\right) + id_z K_0\left(\frac{\omega_0 r_\perp}{v\gamma}\right)\frac{1}{\gamma}\right). \tag{16}$$

Analyzing the commutation relation between the TLS ladder operators shows that the general form of the $S$ matrix is given by:

$$S = e^{-i(Gb\sigma_+ + G^*b\sigma_-) - iK\sigma_z}, \tag{17}$$

where $G$ is getting contributions from the odd orders of the Magnus expansion and $K$ from the even orders. This can be readily seen from the commutation relations of the Pauli operators, e.g., $[\sigma_+, \sigma_-] = \sigma_z$, $[\sigma_z, \sigma_\pm] \propto \sigma_\pm$. This leads to the cyclical behavior in the operators $\Omega_n$ appearing in the Magnus interaction. To estimate the efficiency of using the expansion we need to estimate integrals of the form $\frac{1}{\hbar^n n!}\int dt_1 \ldots dt_n V_I^n(z+vt)$. To get an estimate we look at the electric field of a point charge in the non-relativistic limit:

$$V_I \approx \frac{edr_\perp}{4\pi\varepsilon_0(r_\perp^2 + (z+vt)^2)^{\frac{3}{2}}} = \frac{ed}{4\pi\varepsilon_0 r_\perp^2} \cdot \frac{1}{\left(1+\frac{(z+vt)^2}{r_\perp^2}\right)^{\frac{3}{2}}} = \frac{ed}{4\pi\varepsilon_0 r_\perp^2} \cdot V_I^{\text{dimensionless}}. \tag{18}$$

To make the integral dimensionless we change variables $x = \frac{z+vt}{r_\perp}$, $dt = \frac{dx}{r_\perp v}$.

$$\Omega_n \propto \frac{1}{n!}\left(\frac{ed}{4\pi\hbar v \varepsilon_0 r_\perp}\right)^n \cdot [\text{dimensionless integrals}] \approx \frac{(3 \cdot 10^{-3})^n}{n!} \cdot [\text{dimensionless integrals}], \quad (19)$$

where we took $er_\perp \approx d$ and $v \approx 10^8$ m/s. We see that in general, each order in the Magnus expansion will be 3 orders of magnitude smaller which already justifies our approximation of neglecting $K$. Furthermore, because the dimensionless integrals are oscillatory, and each higher-order term oscillates faster, the discrepancy between higher-order terms in the Magnus expansion should be even larger than expected from a naïve scaling based on Eq. (19). We have corroborated this argument with numerical estimates of the higher-order terms for realistic parameters.

The parameter $g$ is a dimensionless parameter that quantifies the interaction strength. It is useful to define $g \equiv |g|e^{i\phi_g}$, where $|g|$ the magnitude of the interaction strength and $\phi_g$ its phase. This coupling, as can be seen from Eq. (16) is electron-velocity-dependent. To get an estimate of the velocity of the free electron in order to get the strongest interaction possible, we consider atomic dipoles oriented along $x$ and along $z$. We will work in the non-relativistic limit (as we will see the strongest interaction is achieved at velocities well below the speed of light, for which $\gamma_\varepsilon \approx 1$). For a dipole along the $x$ axis, we have:

$$g = -\frac{ie\omega_0 d_x}{2\pi\varepsilon\hbar v^2} K_1\left(\frac{\omega_0 r_\perp}{v}\right) = \frac{a}{v^2} K_1\left(\frac{b}{v}\right), \quad (20)$$

$$\frac{\partial g}{\partial v} = -\frac{2a}{v^3} K_1\left(\frac{b}{v}\right) - \frac{ab}{v^4} K_1'\left(\frac{b}{v}\right) = 0 \Rightarrow -\frac{b}{v} K_1'\left(\frac{b}{v}\right) - K_1\left(\frac{b}{v}\right) = 0. \quad (21)$$

This can be evaluated numerically and results in, $v_{opt} \approx \frac{b}{1.33} = \frac{r_\perp \omega_0}{1.33}$. The same analysis can be performed on a dipole pointing only along the $z$ axis and results in $v_{opt} \approx \frac{b}{1.55} = \frac{r_\perp \omega_0}{1.55}$. So, the optimal velocity depends on the spatial structure of the TLS and will be within the range:

$$\frac{r_\perp \omega_0}{1.55} < v_{opt} < \frac{r_\perp \omega_0}{1.33}. \tag{22}$$

Typically, in the optical range $\hbar\omega_0$ will be few electron volts and $r_\perp$ will be few nanometers. We can write for simplicity $\omega_0 = a \cdot \frac{\text{eV}}{\hbar}, r_\perp = b \cdot \text{nm}$. It follows then that the optimal velocity of the free electron is governed by:

$$a \cdot b \cdot 0.32\% \, c < v_{opt} < a \cdot b \cdot 0.38\% \, c. \tag{23}$$

For realistic parameters, the optimal velocity will always be a few percent of the speed of light and so the approximation $\gamma = 1$ is well justified.

Similar to the traditional PINEM analysis, it is convenient to work in the regime where the electron wave function is a coherent superposition of discrete energies with a small gaussian broadening (much smaller than $\hbar\omega_0$). Thus, we may idealize the electron as a truly discrete superposition of energies evenly spaced by the TLS energy. In this notation, it is convenient to define the electron state as $|\psi_e\rangle = \sum_k C_k |E_0 + k\hbar\omega_0\rangle$. In this notation the electron ladder operators are defined as:

$$b|E\rangle = |E - \hbar\omega_0\rangle, \qquad b^+|E\rangle = |E + \hbar\omega_0\rangle. \tag{24}$$

Under all of these developments, both the definition of the ladder operators and the neglect of the $K$-terms, we may finally arrive at the main result of our manuscript, in which the S-matrix is expressed as:

$$S = \cos|g| - i\sin|g| \left(e^{i\phi_g} b\sigma^+ + e^{-i\phi_g} b^+ \sigma^-\right). \tag{25}$$

**Section IV – The evolution of the TLS before the interaction with the electron**

In this section, we discuss the evolution of the state of the TLS before the interaction with the electron. As the interaction with the electron is very brief (fs time scale) compared to typical decoherence time scales, we can consider the electron-TLS interaction as instantaneous compared to the TLS excitation and relaxation times.

Let us suppose that a laser field or any other means of coherent manipulation acts on the atom at time $\tau = 0$. Hence, before the action of the laser field, the atom has the following density matrix on the basis $|g\rangle, |e\rangle$:

$$\rho(\tau) = \begin{pmatrix} 1 & 0 \\ 0 & 0 \end{pmatrix} \quad (\tau < 0), \tag{26}$$

which represents the atom being in the ground state. At time $\tau = 0$ the atom interacts with a laser pulse and is excited to the coherent superposition $|\psi\rangle = a_1|g\rangle + a_2|e\rangle$, with its density matrix written as:

$$\rho(\tau) = \begin{pmatrix} 1 - |a_2|^2 & a_1 a_2^* \\ a_1^* a_2 & |a_2|^2 \end{pmatrix} \quad (\tau = 0). \tag{27}$$

At this point, the atom goes through relaxation and decoherence from the coupling to the environment, as described by Lindblad master equations [5] and the resulting density matrix as a function of delay $\tau$ is given by:

$$\rho(\tau) = \begin{pmatrix} 1 - |a_2|^2 e^{-\frac{\tau}{T_1}} & a_1 a_2^* e^{i\omega_0 \tau} e^{-\frac{\tau}{T_2}} \\ a_1^* a_2 e^{i\omega_0 \tau} e^{-\frac{\tau}{T_2}} & |a_2|^2 e^{-\frac{\tau}{T_1}} \end{pmatrix}, \tag{28}$$

where $T_1$ and $T_2$ are the longitudinal and transverse relaxation times (or as called sometimes relaxation and decoherence times) and typically $T_2 \ll T_1$; $\hbar\omega_0$ is the energy difference between ground and excited states of qubit.

**Section V – TLS interaction with a general electron wavefunction**

In this section, we investigate the resulting EELS spectra of a general electron wave function after the interaction with the TLS. We consider a wave function composed of a superposition of energies separated by $\hbar\omega_0$. A general wave function for the electron can be written as a sum of wavefunctions with different central energies. We take the initial wavefunction as

$$|\psi_{in}\rangle = \sum_n C_n |E_n\rangle \otimes (a|g\rangle + e^{i\phi_a} b|e\rangle), \tag{29}$$

where $a$ and $b$ are positive real numbers satisfying $a^2 + b^2 = 1$ and $|E_n\rangle$ is defined as $|E_0 + n\hbar\omega_0\rangle$. The final wave function $|\psi_{out}\rangle$, is given as $|\psi_{out}\rangle = S|\psi_{in}\rangle$ so that:

$$|\psi_{out}\rangle = \sum_n C_n |E_n\rangle \otimes (a|g\rangle + e^{i\phi_a} b|e\rangle) \cdot \cos|g|$$

$$-i \cdot \sum_n C_{n+1} |E_n\rangle \otimes (e^{i\phi_g} a|e\rangle) \cdot \sin|g|$$

$$-i \cdot \sum_n C_{n-1} |E_n\rangle \otimes (e^{i(\phi_a - \phi_g)} b|g\rangle) \cdot \sin|g|. \tag{30}$$

What we measure eventually in the EELS spectrum is the probability of the electron to be in specific energy, $P_n = |\langle E_n|\psi_{out}\rangle|^2$, where:

$$\langle E_n|\psi_{out}\rangle = |g\rangle \cdot [aC_n \cdot \cos|g| - ie^{i(\phi_a - \phi_g)} bC_{n-1} \cdot \sin|g|]$$

$$+|e\rangle \cdot [e^{i\phi_a} bC_n \cdot \cos|g| - ie^{i\phi_g} aC_{n+1} \cdot \sin|g|]. \tag{31}$$

and its modulus-squared is given as:

$$P_n = |C_n|^2 \cos^2|g| + |bC_{n-1}|^2 \sin^2|g| + |aC_{n+1}|^2 \sin^2|g|$$
$$+ 2\text{Re}\{i\cos|g|\sin|g|\left(C_n C_{n-1}^* abe^{i(\phi_g - \phi_a)} + C_n C_{n+1}^* bae^{-i(\phi_g - \phi_a)}\right)\}. \quad (32)$$

Expressed in terms of density matrix (as defined in Eq. (26)) elements, we get the spectrum for a general atom (not necessarily an atom in coherent superposition):

$$P_n = |C_n|^2 \cos^2|g| + p|C_{n-1}|^2 \sin^2|g| + (1-p)|C_{n+1}|^2 \sin^2|g|$$
$$+ 2Re\{i\cos|g|\sin|g|\left(C_n C_{n-1}^* q e^{i(\phi_g)} + C_n C_{n+1}^* q^* e^{-i(\phi_g)}\right)\}. \quad (33)$$

Calculating the average energy gain for the initially symmetric electron ($|C_n| = |C_{-n}|$):

$$\langle E_{gain}\rangle = \hbar\omega_0 \sin^2|g|(2p-1) + i\hbar\omega_0 \cos|g|\sin|g|\left(qe^{i(\phi_g)}\sum C_n C_{n-1}^* - q^*e^{-i(\phi_g)}\sum C_n C_{n+1}^*\right). \quad (34)$$

The expression $\sum C_n C_{n-1}^*$ is the expectation value of the ladder operator $b$. Approximating for small $g$ is is convenient to write:

$$\langle E_{gain}\rangle = \hbar\omega_0|g|^2(2p-1) + i\hbar\omega_0|g|\left(qe^{i(\phi_g)}\langle b\rangle - q^*e^{-i(\phi_g)}\langle b^+\rangle\right). \quad (35)$$

The second term comes from the interference between the different electron energies (as it contains the off-diagonal elements of the density matrix). This term decays as the system goes through decoherence. For example, we can look at time $\tau$ for which $\tau \ll T_1$ and assume that $T_2 \ll T_1$, and then the decoherence will be expressed as a decay of the interference term:

$$P_n(\tau) = |C_n|^2 \cos^2|g| + |bC_{n-1}|^2 \sin^2|g| + |aC_{n+1}|^2 \sin^2|g|$$
$$+ 2Re\{i\cos|g|\sin|g|\left(C_n C_{n-1}^* ab^* e^{i\phi} + C_n C_{n+1}^* ba^* e^{-i\phi}\right)\}e^{-\frac{\tau}{T_2}}. \quad (36)$$

We consider two simple configurations of electron states. First, we consider a unshaped electron, for which $C_n = \delta_{n,0}$. This corresponds to the case in many standard EELS and CL microscopy experiments. The electron can either gain/loss energy due to the interaction with the TLS or stay unchanged:

$$P_{-1} = (1 - b^2)\sin^2|g|, P_0 = \cos^2|g|, P_1 = b^2\sin^2|g|, \tag{37}$$

From the EELS spectrum, we can conduct the population statistic of the qubit, this term has only contributions from incoherent terms as the electron cannot undergo interference with itself, and so when the system will go through relaxation the solution will be:

$$P_{-1} = \left(1 - b^2 e^{-\frac{\tau}{T_1}}\right)\sin^2|g|, P_0 = \cos^2|g|, P_1 = b^2\sin^2|g|e^{-\frac{\tau}{T_1}}. \tag{38}$$

By making a repetitive measurement of interaction like that, we can get the EELS spectrum for different time delays $\tau$ and extract information such as energy gaps and lifetimes of atoms, but it does not include any information about the coherent structure of the probed system. This measurement scheme is presented in the main text Fig. 2. In order to observe coherent properties, we will need to use an electron with multiple energy levels distanced from each other by the energy gap of the atomic system. Different energies of the electron interfere during the interaction and create observable changes in the EELS spectrum. The simplest case to check will be an electron with 2 energies distanced $\hbar\omega_0$ with a defined phase between them, such an electron can be created approximately by using a PINEM interaction. It is important to mention that it is not necessary to use exactly the electron that has specifically 2 energies. All that is needed in order to observe quantum interference is that the electron will have at least 2 energies distanced by the energy of the qubit transition $\hbar\omega_0$. It is however the simplest and the most elegant to demonstrate the concept using a duo energetic electron.

The wave function of such an electron can be written as:

$$|\psi_e\rangle = \frac{1}{\sqrt{2}}\left(\left|E_0 - \frac{1}{2}\hbar\omega_0\right\rangle + e^{i\phi_e}\left|E_0 + \frac{1}{2}\hbar\omega_0\right\rangle\right). \tag{39}$$

The resulting EELS spectra will contain 4 peaks, and will be:

$$\begin{cases} P_{-\frac{1}{2}} = \frac{1}{2}\cos^2|g| + \frac{1}{2}a^2\sin^2|g| - \cos|g|\sin|g|\,ab\cdot\sin(\phi_a - \phi_e - \phi_g) \\ P_{\frac{1}{2}} = \frac{1}{2}\cos^2|g| + \frac{1}{2}|b|^2\sin^2|g| + \cos|g|\sin|g|\,ab\cdot\sin(\phi_a - \phi_e - \phi_g) \\ P_{-\frac{3}{2}} = \frac{a^2}{2}\sin^2|g| \\ P_{\frac{3}{2}} = \frac{b^2}{2}\sin^2|g| \end{cases}, \tag{40}$$

Since the coupling is typically weak, we can approximate and get

$$\begin{cases} P_- = \frac{1}{2} - |g||ab|\sin(\phi_a - \phi_e - \phi_g) \\ P_+ = \frac{1}{2} + |g||ab|\sin(\phi_a - \phi_e - \phi_g) \end{cases}, \tag{41}$$

This is an interesting expression, as the phase $\phi_e$ is controlled by the shaped electron, and the phase $\phi_g$ depends on the internal structure of the TLS, such as the size of the dipole in each direction, and the phase $\phi_a$ depends on the coherent superposition of the TLS. That way, we can use a precise phase for the electron wave function in order to measure the phase difference of the TLS states. Alternatively, we can measure the relative size between the different dipole components $d_x$ and $d_z$ by measuring $\phi_g$. Another notable property of this expression is that the contrast in the resulting EELS signal is proportional to $|g|$, as opposed to $|g|^2$.

Eq. (35) shows us that in general, we can get electron energy gain and loss that is proportional to $|g|$ rather than $|g|^2$ when $\langle b\rangle \neq 0$, presenting how the reliance on quantum interference is used

here to not only extract information about the coherent phases of the TLS but also to intensify the energy transfer between the electron and the TLS. This kind of enhanced interaction will lead eventually to an enhanced CL signal, as the electron transfers this energy to the excitation of the TLS which will result eventually in a stronger emission of radiation. It is important to mention that a PINEM-generated electron without free space propagation or other amplitude manipulation will not have an energy gain/loss proportional to $|g|$. To show this, we recall that the wavefunction of PINEM modulated electron [6] with laser frequency equal to $\omega = \omega_0/l$:

$$|\psi_e\rangle = \sum_n e^{i\phi n} J_n(2|g^{PINEM}|)|E_0 + \hbar n\omega\rangle. \qquad (42)$$

The expectation value of the $b$ operator will always be zero according to the Bessel Functions identity:

$$\sum_n J_n(x) J_{n-l}(x) = 0. \qquad (43)$$

Nevertheless, the EELS signal still contains features proportional to $|g|$, such as in the height difference between EELS peaks. For the average gain to change, or to get an enhancement of the energy transfer, the initial electron must be shaped by amplitude modulation and not only phase modulation. This can be achieved having a distance of free space propagation after a PINEM interaction. The PINEM only induces phase modulation, but then the electron dispersion in free space transforms the phase modulation into amplitude modulations. Similar enhancements, based on the modulation of free electrons, were proposed for the first time using a semiclassical analysis in [4]. We show how our quantum description of the interaction conforms to part of the results of the semiclassical theory [4] and also generalizes related work on the quantum klystron [7]. The semiclassical formalism used in these papers can describe the enhancement of the interaction with qubits due to the phase modulation of the electrons but cannot describe the resulting EELS spectrum as it relies on quantum interference, which is crucial for quantum measurements. Another way to enhance the CL due to the interaction with free electrons is described in Section VI.

According to Eq. (16), we can estimate the interaction strength for perovskites. We consider a typical transition dipole moment of $d \approx 288$ D, excited state energy $\hbar\omega_0 \approx 1\text{eV}$, electron speed of 7% the speed of light, and a distance from the qubit $r_\perp = 6$ nm. For these parameters, the interaction strength is approximately $|g| \approx 0.1$. The typical dipole of the molecules is usually much smaller. For example, the dipole moment of a water molecule is $d \approx 2.0\ D$. In this case, the interaction constant equals to: $|g| \approx 0.7 \cdot 10^{-3}$.

The standard deviation of the EELS features after $N$ measurements can be estimated as $\frac{1}{\sqrt{N}}$. It means that the relative mistake of the interference measurements (where the measured result is $\sim |g|$) approximately equals $\frac{1}{g\sqrt{N}} \cdot 100\%$. To have a standard deviation of 1% of the result, in the case of the strongest interaction of $|g| = 0.1$, we need approximately $10^6$ repetitions, i.e. million electrons.

**Section VI – Observing of qubits emitting radiation superradiantly**

In this section we consider weak interaction ($|g| \ll 1$) of a free electron with a group of $N$ non-interacting qubits close to each other, and see how their superradiant behavior can be observed through the EELS signal. In this case, the scattering matrix $S$ will have the following form:

$$S = \prod_i S_i = \prod_i e^{-i(g\sigma_+^i b + g^* \sigma_-^i b^+)}, \tag{44}$$

where $\sigma_{\pm}^i$ are Pauli matrices for $i^{th}$ qubit. If the qubits are prepared in the collective sate $|\psi_A\rangle$ and them interacts with a monoenergetic electron $|E_0\rangle$, in case of weak interaction strength ($|g| \ll 1$) we can approximate the resulting final state $|\Psi_f\rangle$ to second order in $|g|$:

$$|\Psi_f\rangle = S|E_0\rangle|\psi_A\rangle \approx$$

$$\approx (1 - N|g|^2)|E_0\rangle|\psi_A\rangle - i\left[\sum_i g\sigma_+^i |E_0 - \hbar\omega_0\rangle|\psi_A\rangle + \sum_i g^*\sigma_-^i |E_0 + \hbar\omega_0\rangle|\psi_A\rangle\right]. \quad (45)$$

Then the resulting spectrum has the following form:

$$\begin{cases} P_0 = 1 - N|g|^2 \\ P_+ = |g^2|\sum_i \langle\psi_A|\sigma_-^i|\psi_A\rangle = |g|^2 \cdot \left(\frac{2\langle E_{qubits}\rangle}{\hbar\omega_0} + \frac{N}{2}\right) \\ P_- = |g^2|\sum_i \langle\psi_A|\sigma_+^i|\psi_A\rangle = -|g|^2 \cdot \left(\frac{2\langle E_{qubits}\rangle}{\hbar\omega_0} - \frac{N}{2}\right) \end{cases}, \quad (46)$$

where $P_0$ is the probability that the energy of the electron will not be changed, $P_+$ is the probability that an electron's energy will increase on $\hbar\omega_0$ and $P_-$ is the probability that an electron's energy will decrease on $\hbar\omega_0$. Then, the average energy gain equals to:

$$\langle E_{\text{gain}}(\tau)\rangle = \hbar\omega_0(\Delta P) = \hbar\omega_0|g|^2 \frac{4\langle E_{\text{qubits}}\rangle}{\hbar\omega_0}, \quad (47)$$

where $\langle E_{\text{qubits}}\rangle$ is the average energy of the superradiant qubits. The intensity of superradiant emission is connected with $\langle E_{\text{qubits}}\rangle$ [8]:

$$I_{\text{superradiant}}(\tau) = -\frac{d\langle E_{\text{qubits}}(\tau)\rangle}{dt}. \quad (48)$$

Hence, measuring the gain for different time delays and using Eq. (47) we can reconstruct the superradiant emission of the qubits.

Eq. (45) show us that the electron states with energy gain/loss are entangled to qubit states raised *symmetrically*, this tells us that even if we didn't pre-excite the qubits as suggested in the measurement scheme the interaction with the electron results in an (at least partly) symmetric state, such state as we know, given that the qubit-qubit interactions are weak enough and the qubits are close enough, will emit superradiantly [8]. This kind of superradiant emission is a form of a coherent enhancing of the CL signal.

**Section VII – Extracting the local density of photonic states (LDOS)**

In this section, we discuss how LDOS can be extracted from the interaction with free electrons. In cases of qubits with weak non-radiative relaxation, $T_1$ is directly related to the qubit's spontaneous emission rate $\gamma = 1/T_1$, and is influenced by its optical environment – the local density of optical states (LDOS). An emitter at a location $\boldsymbol{r}_0$ and oriented along direction $z$ has [9]:

$$\gamma = \frac{\omega_0}{\hbar\varepsilon_0}|\boldsymbol{d}|^2 \rho_z(\boldsymbol{r}_0,\omega_0) = \frac{4\omega_0^2}{\pi\hbar\varepsilon_0 c^2}|\boldsymbol{d}|^2 \mathrm{Im}\{G_{zz}(\boldsymbol{r}_0,\boldsymbol{r}_0,\omega)\}, \qquad (49)$$

where $G_{zz}(\boldsymbol{r}_0,\boldsymbol{r}_0,\omega)$ is the $zz$ component of the Green function and $\rho_z(\boldsymbol{r}_0,\omega_0)$ is the LDOS. Typically, $\gamma$ is measured by optically exciting a group of emitters in an area and measuring the resulting emission as a function of time [9]. Indirect measurement is through the spectrum [9], where the broadening of the peak depends on the lifetime (competing with other processes of incoherent broadening). Those techniques are limited by their spatial resolution due to the optical wavelength and often require many emitters to collect sufficient signal.

Advances in CL enable measuring the LDOS with deep sub-wavelength resolution using an electron probe. More recent advances use time-resolved CL to achieve a direct measurement of $T_1$ at such deep-subwavelength resolutions [11, 12]. Using shaped-electron–qubit interactions poses an alternative way of measuring the LDOS. Its relative advantage is the femtosecond time resolution that arises from the durations of the electron pulses and laser pulses in our scheme.

**Section VIII – Measurement of a qubit state**

In this section, we discuss further the measurement scheme proposed in Section III in the main text on how to measure the complete qubit state on the Bloch-sphere using coherently shaped free electrons. Eq. (40) gives us the resulting EELS spectrum of a duo energetic electron interacting with a qubit in general coherent superposition. The height difference between the two main peaks is given by:

$$\Delta P = (P_+ + P_-)|g|\sin(\theta_a)\sin(\Phi), \qquad (50)$$

where $\Phi = \phi_a - \phi_e - \phi_g$ and $\theta_a, \phi_a$ are the angles describing the qubit location on the Bloch-sphere (Fig. 3 on the main text). By scanning on different electron phases one can find the phase $\phi_e$ for which $\Delta P$ is maximal, in this phase the condition $\sin(\Phi) = 1$ is achieved and $\phi_a$ is extracted. Then by measuring $\Delta P$ we conclude $|g|\sin(\theta_a)$. This method however gives an inconclusive answer about the qubit state as the function $\sin(\theta_a)$ is multivalued in the range $[0, \pi]$ leaving an ambiguity on whether the qubit is in the upper or lower half of the Bloch-sphere. To solve that one will need to look at the other side lobes in the EELS spectrum which corresponds to the energies $E_+ + \hbar\omega_0$ (gain lobe) and $E_- - \hbar\omega_0$ (loss lobe) as they are not affected by quantum interference. If the gain lobe is higher, then the qubit was in the lower half of the Bloch-sphere (meaning that without quantum interference it will mostly "give" energy to the electron) and otherwise, on the upper half. The drawback of this method is that the height of the side lobes is proportional to $|g|^2$ increasing the required sensitivity of the EELS measurement or alternatively increasing the required number of repetitions in order to get a measurement. A possible way to pass this drawback will be to "flip" the qubit 90 degrees around the x-axis using conventional coherent control methods and then measure again with the same method, giving us again probability difference that goes like $|g|$.